\documentclass[12pt,draftclsnofoot,journal,onecolumn]{IEEEtran}

\usepackage[dvips]{color}
\usepackage{epsf}
\usepackage{times}
\usepackage{epsfig}
\usepackage{graphicx}
\usepackage{url}
\usepackage{amsmath}
\usepackage{amssymb}
\usepackage{amsxtra}
\usepackage{algorithmic}
\usepackage{multirow}
\usepackage{cite}
\usepackage{enumerate}
\usepackage{subfigure}
\usepackage[ruled]{algorithm}

\newtheorem{property}{\bf Property}
\newtheorem{definition}{\bf Definition}

\newlength{\aligntop}
\newlength{\alignbot}
\makeatletter
\renewenvironment{align}{%
  \vspace{\aligntop}
  \start@align\@ne\st@rredfalse\m@ne
}{%
  \math@cr \black@\totwidth@
  \egroup
  \ifingather@
    \restorealignstate@
    \egroup
    \nonumber
    \ifnum0=`{\fi\iffalse}\fi
  \else
    $$%
  \fi
  \ignorespacesafterend%
  \vspace{\alignbot}\par\noindent
}

\IEEEoverridecommandlockouts
\begin{document}

\title{Improving Macrocell - Small Cell Coexistence \\ through Adaptive Interference Draining}
\author{\authorblockN{Francesco Pantisano$^{1,2}$, Mehdi Bennis$^{1}$, Walid Saad$^{3}$, M{\'e}rouane Debbah$^{4}$ \\ and Matti Latva-aho$^1$\\} \bigskip \authorblockA{\footnotesize
$^\textbf{1}$ Centre for Wireless Communications - CWC, University of Oulu, 90570, Finland, email: \url{{fpantisa,bennis,matla}@ee.oulu.fi}\\
$^\textbf{2}$ Dipartimento di Elettronica e dell' Informazione - University of Bologna, 40135, Italy, email:\url{{francesco.pantisano}@unibo.it}\\
$^\textbf{3}$ Electrical and Computer Engineering Department, University of Miami, Coral Gables, FL, USA, \url{walid@miami.edu}.\\
$^\textbf{4}$ Alcatel-Lucent Chair in Flexible Radio,  SUP{\'E}LEC, Gif-sur-Yvette, France email: \url{merouane.debbah@supelec.fr}
 }%
   \thanks{The authors would like to thank the Finnish foundation for technology and innovation promotion, the Nokia Foundation, the Riitta and Jorma J. Takanen Foundation, Elektrobit and Nokia Siemens Networks for supporting this work. }}
\date{}
\maketitle

\thispagestyle{empty}

\begin{abstract}
The deployment of underlay small base stations~(SBSs) is expected to significantly boost the spectrum efficiency and the coverage of next-generation cellular networks. However, the coexistence of SBSs underlaid to an existing macro-cellular network faces important challenges, notably in terms of spectrum sharing and interference management. In this paper, we propose a novel game-theoretic model that enables the SBSs to optimize their transmission rates by making decisions on the resource occupation \emph{jointly} in the frequency and spatial domains. This procedure, known as \emph{interference draining}, is performed among cooperative SBSs and allows to drastically reduce the interference experienced by both macro- and small cell users. At the macrocell side, we consider a modified water-filling policy for the power allocation that allows each macrocell user~(MUE) to focus the transmissions on the degrees of freedom over which the MUE experiences the best channel and interference conditions. This approach not only represents an effective way to decrease the received interference at the MUEs but also grants the SBSs tier additional transmission opportunities and allows for a more agile interference management. Simulation results show that the proposed approach yields significant gains at both macrocell and small cell tiers, in terms of average achievable rate per user, reaching up to $37\%$, relative to the non-cooperative case, for a network with $150$ MUEs and $200$ SBSs.
\end{abstract}

\newpage

\section{Introduction}

The use of underlaid small cell base stations~(SBSs) has been proposed in the upcoming wireless standards, such as Long-Term Evolution Advanced\cite{3GPP1}, so as to increase the spectral efficiency and improve the indoor coverage\cite{LTE-A2}. SBSs are low-cost, low-power, base stations that can be deployed either outdoor by the operator (e.g., picocells, microcells, or metrocells) or indoor by end users (e.g., femtocells) so as to boost the capacity of wireless systems by reducing the distance between users and their serving stations. Since, in an underlay spectrum access, the SBSs opportunistically reuse the macrocell spectrum, the interference has been identified as the main limiting factor for the macrocell-small cell coexistence \cite{5876497,DP1}. In this context, two major interference components are identified: the interference brought from the SBSs to the macrocell users~(MUEs), and the interference among different SBSs, which are respectively referred to as cross-tier and co-tier interference. While co-tier interference is a major challenge in all SBS deployments, cross-tier interference is particularly sever in outdoor small cell deployment, such as operator-deployed picocells \cite{SC1}. Moreover, as the SBSs can only access the spectral resources which are under-utilized by the macrocell tier, the small cells are not provided any guarantees in terms of transmission opportunities or quality of service~(QoS) requirements. Hence, developing efficient interference management and spectrum access policies is of utmost importance for achieving the performance foreseen for small cell deployments~\cite{3GPP2}.

Recently, significant research efforts have been dedicated to the study of macrocell- small cell coexistence, by relying on the SBSs' self-organization capabilities. Notably, dynamic spectrum access \cite{5876497,4720238,5375679,niyato1,joint1}, interference coordination \cite{5876497,6008531,6190344} and power control \cite{1626432,Andrews2,Hwang,sing1} have been proposed for managing interference by exploiting frequency carriers or time slots that are under-utilized in the macrocell tier.
In this context, interference alignment~(IA) has been proposed as a linear coding technique that can virtually guarantee interference-free transmissions by exploiting the spatial directions of the multiple-input multiple-output~(MIMO) interference channel, referred to as degrees of freedom, hereinafter~\cite{SJ1,tse1,SJ4}. As IA techniques only use half of the available transmission opportunities, the opportunistic exploitation of the spatial degrees, proposed in the IA concept, has been recently extended to incorporate the frequency dimension. By doing so, one can improve the transmission rates of the small cells by exploring new transmission opportunities (i.e., degrees of freedom) in both the frequency and space domains \cite{SP1,amir,Heath}. It thus becomes possible for the small cells to enable interference-free communications by leveraging the spatial and frequency precoding. Moreover, the opportunistic IA solution can be combined with other interference management techniques, such as successive interference cancellation~\cite{6096932} or zero forcing equalization \cite{6302137}. In order to perform opportunistic IA, the complete knowledge of the full channel state information~(CSI) is required and the transmitters are required to cooperate for the joint design of the precoding matrices. However, IA techniques are limited by the fact that IA solutions only exist for certain problem dimensions, which are given by the number of antennas and the degrees of freedom \cite{SJ4,SJ5}. Hence, in practice, an IA scheme one can only suppress a limited number of interfering signals, since the number of antennas (especially at the receiver side) is limited. Moreover, using half of the spatial degrees of freedom further reduces the already scarce transmission opportunities of underlaid small cell networks\cite{SJ4}.


To overcome these IA limitations, the concept of \emph{interference draining}~(ID)\cite{drain} has been recently introduced with the purpose of reusing the degrees of freedom unused by the macrocell, while controlling the interference brought to the macrocell tier. In essence, ID is an extension of the IA solution to the case of shared spectrum deployments. Some of the conditions on the mutual alignment of the interfering signals are relaxed, and a margin of interfering power is allowed at each receiver. In addition, such an approach has full reuse of the degrees of freedom in both space and frequency domains, provided that the interference constraints are verified. In \cite{drain}, an opportunistic technique for interference-limited networks is presented to enable the interference draining in the space and time domain and increase the number of secondary users in the system. In \cite{6214446}, the extension of IA to the time domain is combined with a partial alignment technique for data rate enhancement of secondary networks. Note that both interference draining and interference alignment involve operations which are jointly performed by mutual interferers, which, however, are uncoordinated. Therefore, it is clear that there is a need for novel cooperative strategies at the SBS level aimed at a dynamic reuse of the macrocell spatial and frequency resources, as recently suggested in ~\cite{6157448,6008531,Zhuli,5936229,SP1,noncoop1}.

The main scope of this paper is to jointly address the co-tier and the cross-tier interference management in the downlink of an underlay macrocell-small cell network. Due to the highly dynamic changes in the small cell tier (e.g., SBSs turning on/off, dynamic user arrivals), the optimization of macro base station~(MBS) transmissions accounting for the bursty interference generated by the SBSs is a very complex task. In this respect, we first show that, when the MBS performs an independent and interference-unaware power allocation \cite[Section 7.1.1]{tsebook}, the achievable downlink data rates are strongly affected by the small cell interference. This effect becomes more acute when the MBS transmits several signal streams over the channels degrees of freedom, and for large small cell tiers.
Hence, we propose that the SBSs entirely manage the co-tier and cross-tier interference by \emph{cooperatively} using an interference draining technique. Unlike existing work which addresses static cooperation among the secondary nodes, i.e., implicit cooperation among mutual interferers \cite{SJ1,drain,6214446}, our proposed approach allows the SBSs to autonomously decide on \emph{when} to cooperate and with \emph{whom}, based on their self-organizing capabilities. One of the key advantages of the proposed solution is that, by opportunistically reusing the channel degrees of freedom, the small cells suppress the mutual interference, while still satisfying a minimum QoS requirement at the nearby MUEs. Note that the proposed approach does not require direct coordination from the MBS, which, in turn, can optimize its spectrum access and power allocation independently of underlaid small cell transmissions.

In a second phase, we investigate the benefits of low-overhead cooperation among MUEs and SBSs. Here, we propose that the MUEs alleviate the interference produced by the SBSs by adjusting the MBS' power allocation and focusing the transmission only on the degrees of freedom experiencing the best channel and interference conditions. In other words, we propose that an interference-limited MUE can deliberately release those degrees of freedom, if this allows for a reduction of the interference over the degrees of freedom which remain in its use. This approach, which is akin to the modified water-filling policy \cite{ioannis}, creates new transmission opportunities for the nearby SBSs.

We formulate the problem of macrocell- small cell coexistence as a coalitional game in which the SBSs and the MUEs are the players. By deciding to cooperate, the players increase their own utility in terms of achievable data rate, while accounting for both co-tier and cross-tier interference constraints. We show that, due to the mutual interference, the utility achieved by any player is affected by the cooperative behavior of the other players in the network. As a result, the proposed small cell coalitional game is in partition form, which is a class of coalitional games significantly different than classical characteristic form games, widely studied in wireless networks \cite{5936229,noncoop1}. We solve the considered game through the concept of a recursive core \cite{K01}, a key solution concept for coalitional games in partition form.

\noindent In summary, our key contributions are the following:

\begin{itemize}
       \item We design a framework in which the small cells are underlaid to a macrocell network and reuse the macrocell degrees of freedom in the space and frequency domains.
       \item We propose a small cell interference mitigation solution based on ID, in which the SBSs autonomously manage the interference brought to the other cooperative SBSs and the MUEs in their vicinity.
       \item The proposed cooperative approach takes advantage of the different nature of co-tier and cross-tier interferences. On the one hand, underlaid SBSs are mostly limited by resource availability and co-tier interference. Hence, it is beneficial to mitigate the interference while keeping the small cell transmissions confined in a limited frequency band different from the one used by the nearby MUEs. On the other hand, the cooperative SBSs' transmissions are required to satisfy the QoS requirements of the MUEs in proximity.
       \item We model a small cell cooperative behavior using a game theoretical approach, by formulating a coalitional game in which MUEs and SBSs are the players. The benefits from cooperation are quantified in terms of improved achievable data rates.
       \item By leveraging the solution of modified water-filling power allocation, we propose a protocol for \emph{implicit} cross-tier cooperation which does not involve direct coordination between the macrocell and the small cell tiers. The proposed protocol enables the MUEs to capitalize on the release of some degrees of freedom with the reduction of the received interference.
       \item We present a distributed coalition formation algorithm through which MUEs and SBSs take autonomous decisions on the selection of a cooperative strategy and reach a stable solution.
\end{itemize}

The rest of this paper is organized as follows. In Section~II, we describe the considered system model and analyze the limitations of the non-cooperative approach. In Section~III we describe the cooperative behavior of MUEs and SBSs for mutual interference management. In Section~IV we model the cooperative framework as a coalitional game, discuss its properties and provide a distributed algorithm for performing coalition formation. Numerical results are discussed in Section~V and finally conclusions are drawn in Section~VI.

\noindent \textbf{Notations}: In the rest of the paper, The $\log$ refers to $\log_2$. Bold uppercase letters (e.g., $\textbf{[A]}_{a \times b}$) denote matrices with $a$ rows and $b$ columns, bold lowercase letters (e.g., $\textbf{a}$) denote column vectors and normal letters (e.g., $a$) denote scalars. The identity matrix is denoted by $\textbf{1}$. The operator $\left \|\cdot \right \|_F$ denotes the Frobenius norm. $\mathbb{C}$ represents the set of complex numbers and  $(\cdot)^{\dag}$ denotes the Hermitian transpose operator.


\bigskip
\section{System Model}\label{sec:sm}
Consider the \emph{downlink} of a single macrocell wireless network in which $K$ SBSs are underlaid to a tier of $N$ MUEs. Both the MBS and the SBSs use an orthogonal frequency division multiple access~(OFDMA) technique over the shared macrocell spectrum. The macrocell bandwidth is divided into non-overlapping frequency subchannels, denoted by the set $\Phi$, and each subchannel represents the unitary spectral resource which can be assigned to each signal stream. Let $\mathcal{K}= \left \{ 1,...,K \right \}$ and $\mathcal{N}= \left \{ 1,...,N \right \}$ denote the sets of the SBSs and the MUEs, respectively. Every SBS $k \in \mathcal{K}$ services $L_k$ small cell users~(SUEs), denoted by  $\mathcal{L}_k= \left \{ 1,...,L_k \right \}$, over $|\Phi_k|$ subchannels, in which $\Phi_k \subset \Phi $ denotes the set of such selected subchannels. Similarly, the MBS allocates a set of subchannels $\Phi_n$ to each MUE $n$, and thus, due to the unitary frequency reuse, $\bigcup_{k \in \mathcal{K}} \Phi_k \cup \: \bigcup_{n \in \mathcal{N}} \Phi_n \subseteq \Phi$. The MBS and SBSs are respectively equipped with $A_n$ and $A_k$ transmitting antennas, while the MUEs and SUEs are both equipped with $B$ receiving antennas. The MBS and each of the SBSs respectively utilize the linear precoding matrices $\textbf{V}_n \in \mathbb{C}^{B \times d_n}$ and $\textbf{V}_{k} \in \mathbb{C}^{B \times d_k}$ to transmit $d_n \leq A_n$ and $d_k \leq A_k$ streams to the corresponding receivers. For the sake of simplicity, we consider that each MUE $n$ is assigned $|\Phi_n|=1$ frequency subchannel, over which $d_n$ signal streams are modulated and transmitted \footnote{Nevertheless, the proposed solution can accommodate multiple subchannel allocation schemes in the macrocell tier, without loss of generality.}. As a result, for each time instant, the discrete received signal at the MUE $n$ is given by:

\begin{equation}\label{eq:rxs}
\textbf{y}_n = \textbf{H}_{0n}\textbf{V}_n\, \textbf{s}_n + \sum_{k \in \mathcal{K}_{\Phi_n}} \textbf{H}_{kn}\textbf{V}_k\,\textbf{s}_k+ \textbf{z}_n,
\end{equation}

\noindent where $\mathcal{K}_{\Phi_n}=\{k \in \mathcal{K} : \Phi_k \cap \Phi_n \neq \emptyset \}$ denotes the subset of the SBSs which are interfering with the macrocell transmission over $\Phi_n$. $[\textbf{H}_{0n}]_{A_n \times B}$ and $[\textbf{H}_{kn}]_{A_k \times B}$ are complex matrices corresponding to the MIMO channels coefficients between the MBS denoted by the subscript $0$ and MUE $n$, and the interfering link between SBS $k$ and MUE $n$, respectively. $\mathbf{s}_n \in \mathbb{C}^{d_n \times 1} $ represents the $d_n$-dimensional signal transmitted to the MUE $n$. In addition, $d_n$ denotes the degrees of freedom of the transmitter-receiver pair (i.e., the number of transmitted signal streams), for the transmitted message. Similarly, $\mathbf{s}_{k} \in \mathbb{C}^{d_k \times 1} $ is the $d_k$-dimensional signal pertaining to SBS $k \in \mathcal{K}_{\Phi_n}$ (that is interfering). Further, $\mathbf{z}_n$ represents the noise vector at MUE $n$ which is considered as a zero mean circularly symmetric additive white Gaussian noise~(AWGN) vector with variance $\sigma^2$. Both transmitted signals $\textbf{s}_n$ and $\textbf{s}_k$ are limited by the respective power constraints $P_{max}^{n}$ and $P_{max}^{k}$ over their signal components: $\sum_d^{d_n} P_d^{n} \leq P_{max}^{n}$, $\sum_d^{d_k} P_d^{k} \leq P_{max}^{k}$, where $P_d^{n}, P_d^{k}$ are the power of the $d$-th signal stream from the MBS to MUE $n$ or from SBS $k$ to each of its SUEs\footnote{Clearly, $P_d^{n}=0$ ($P_d^{k}=0$) if the MBS (SBS $k$) is not transmitting on the $d$-th degree of freedom of the wireless channel.}. In such a setting, we consider that the MBS optimizes its transmissions by neglecting the existence of the small cell tier, and thus, it does not account for the interference generated by the SBSs. In turn, the SBSs are required to adapt their transmission schemes to the current macrocell spectrum allocation so as to control the interference brought to the nearby MUEs and the other SUEs. Through this assumption, which is common in non-cooperative networks \cite{noncoop1,amir,wu}, the macrocell tier can optimize its own transmissions while remaining oblivious of the underlaid small cell transmissions, which allows for higher scalability of the small cell tier. Accordingly, the MBS performs a classical water-filling power allocation over the set of antennas $A_k$ as in \cite[Section 7.1.1]{tsebook}, based the knowledge of the channel realizations $\textbf{H}_{0n}$. Finally, the data rate at MUE $n$ is computed by transforming the MIMO channel to $d_n$ parallel channels, in which one signal stream is transmitted, and can be expressed as \cite{nosrat}:

\begin{equation}\label{eq:ratem}
R_{n}= \sum_{d=1}^{d_n} \log \Bigg(1+ \frac {\gamma_d^{n} / d_n}{\textbf{e}_d \Big( (\textbf{V}_n^{\dag} \textbf{H}_{0n}^{\dag}\textbf{G}_n \textbf{H}_{0n}\textbf{V}_n  )^{-1} + I_n^{\mathcal{K}} \Big) \textbf{e}_d ^{\dag} } \Bigg),
\end{equation}



\noindent where $\gamma_d^{n} = \frac{P_d^{n}}{\sigma^2}$, $\gamma_d^{k} = \frac{P_d^{k}}{\sigma^2}$, $\textbf{e}_d$ is the $d$-th column of $\textbf{1}_{d_n}$ and $\textbf{G}_n = (\textbf{1}_{A_n}- \textbf{b}_n \textbf{b}_n^{\dag})$ denotes the matrix of the projection into the nullspace of the interference subspace of MUE $n$, which is identified by the non-unique basis $\textbf{b}_n$. In addition, $I_n^{\mathcal{K}}=\sum_{k \in \mathcal{K}_{\Phi_n}} \frac{\gamma_d^{k}}{d_k} \textbf{Q}_{n}\textbf{H}_{kn}\textbf{V}_k \textbf{V}_k^{\dag} \textbf{H}_{kn}^{\dag} \textbf{Q}_{n}^{\dag}$ denotes the covariance of the interference brought to MUE $n$ by the co-channel SBSs and $[\textbf{Q}_{n}]_{d_n \times A_n}$ is the respective post-processing matrix at the MUE's side.

From the small cell perspective, spectrum access is carried out in an uncoordinated fashion at each SBS. This implies that, in order to transmit a signal $\mathbf{s}_k \in \mathbb{C}^{d_k \times 1}$, an SBS $k$ selects a set $\Phi_k$ of frequency subchannels, which are potentially affected by both co-tier and cross-tier interference. In this context, a traditional frequency modulation technique (e.g., OFDM) requires $|\Phi_k|=d_k$ subchannels for the signal transmissions. Moreover, using such scheme, each SBS needs to perform additional operations for interference management (e.g., power control \cite{constr2} or bandwidth partitioning \cite{joint1}). In contrast, a spatial coding technique (e.g., interference draining or alignment) allows multiple streams to be transmitted over the same interference-free subchannel, thus it requires $ |\Phi_k| < d_k$ subchannels for transmitting $d_k$ signal streams. As a result, we consider that $1 \leq |\Phi_k| \leq d_k$ for each SBS $k$, as it captures two important features. First, spatial coding transmission techniques increase the spectrum efficiency of an OFDM scheme by enabling $d_k$-dimensional signal transmissions over $ |\Phi_k| \leq d_k$ subchannels. Second, the co-channel interference is avoided through cooperative linear precoding at the transmitter side.

With these considerations in mind, for a transmission from an SBS $k$ to one of its SUEs $i \in \mathcal{L}_k$, the discrete-time received signal at the SUE $i$, at a given time instant, is given by:

\begin{equation}\label{eq:sigf}
\textbf{y}_i = \textbf{H}_{ki}\textbf{V}_k\, \textbf{s}_k + \sum_{j \in \mathcal{K}_{\Phi_k}, \, j \neq k} \textbf{H}_{ji}\textbf{V}_j\,\textbf{s}_j+  \sum_{n \in \mathcal{N}_{\Phi_k}} \textbf{H}_{0i}\textbf{V}_n\,\textbf{s}_n + \textbf{n}_i,
\end{equation}

\noindent where $\mathcal{K}_{\Phi_k} =\{j \in \mathcal{K}, j \neq k : \Phi_j \cap \Phi_k \neq \emptyset \}$, $ \mathcal{N}_{\Phi_k} = \{n \in \mathcal{N} : \Phi_n \cap \Phi_k \neq \emptyset \}$ respectively denote the subsets of SBSs and MUEs whose transmissions are interfering with SUE $i$ over the bandwidth $\Phi_k$. $[\textbf{H}_{ki} ,\textbf{H}_{ji} ]_{A_k \times B}$ respectively denote the complex matrices of the MIMO channels coefficients between SBS $k$ and SUE $i$, and the interfering link between SBS $j$ and SUE $i$, over the used subchannel \footnote{In case of $|\Phi_k| >1$, each of the matrices $[\textbf{H}_{ki} ,\textbf{H}_{ji} ]$ corresponds to one of the used frequency subchannel in $\Phi_k$. Here, we omit the subchannel index for the sake of a simplified notation.}. $\mathbf{s}_j  \in \mathbb{C}^{d_k \times 1} $ denotes the $d_k$-dimensional signals transmitted by SBS $j \in \mathcal{K}_{\Phi_k}$. Finally, the last summation in (\ref{eq:sigf}) represents the interference from the MBS transmitting to its MUE $n$, in which $[\textbf{H}_{0i}]_{A_n \times B}$ is the matrix of the MIMO interference channel between the MBS $0$ and the SUE $i$. With this considerations in mind, we express the rate achieved at each SUE $i \in \mathcal{L}_k$ as \cite{nosrat}:

\begin{equation}\label{eq:ratef}
R_{i}= \sum_{d=1}^{d_k} \log \Bigg(1+ \frac {\gamma_d^{k} / d_k}{\textbf{e}_d \Big( (\textbf{V}_k^{\dag} \textbf{H}_{ki}^{\dag}\textbf{G}_i \textbf{H}_{ki}\textbf{V}_k  )^{-1} + I_i^{\mathcal{K}} + I_i{^\mathcal{N}}  \Big) \textbf{e}_d ^{\dag} } \Bigg),
\end{equation}

\noindent where $[\textbf{Q}_{i}]_{d_i \times A_i}$ is the post-processing matrix at the SUE $i$, $\gamma_d^{k} = \frac{P_d^{k}}{\sigma}$, $\textbf{e}_d$ is the $d$-th column of $\textbf{1}_{d_k}$ and $\textbf{G}_i = (\textbf{1}_{A_k}- \textbf{b}_i \textbf{b}_i^{\dag})$ denotes the matrix of the projection into the nullspace of the interference subspace of SUE $i$, which is identified by the non-unique basis $\textbf{b}_i$. We let $I_i^{\mathcal{K}}= \sum_{j \in \mathcal{K}_{\Phi_k}} \frac{\gamma_d^{j}}{d_j} \textbf{Q}_{i}\textbf{H}_{ji}\textbf{V}_j \textbf{V}_j^{\dag} \textbf{H}_{ji}^{\dag}\textbf{Q}_{i}^{\dag}$ and $I_i^{\mathcal{N}}= \sum_{n \in \mathcal{N}_{\Phi_k}} \frac{\gamma_d^{n}}{d_n} \textbf{Q}_{i}\textbf{H}_{0i}\textbf{V}_n \textbf{V}_n^{\dag} \textbf{H}_{0i}^{\dag}\textbf{Q}_{i}^{\dag}$ denote the covariances of the interfering transmissions from the SBSs and the MBS, respectively.

It can be noted that the performance of the MUEs and SUEs are limited by different factors. While the former are solely limited by the cross-tier interference, the latter face the challenges of the availability of degrees of freedom and the contention with the other uncoordinated SBSs over the transmission opportunities, which incurs severe co-tier interference.

\bigskip
\section{Proposed Cooperative Interference Management Framework}
In this section, we propose two novel cooperative mechanisms of cooperation that enable SBSs to maximize their transmission rate with a constraint on the interference brought to the macrocell tier. We initially propose an interference management scheme which relies on the small cell's self-organization capabilities. Subsequently, we extend the model by including partial cooperation from the MUE side, which, however, requires a limited feedback from the SBSs.

\subsection{Cooperative spatial coding techniques for small cell transmissions}\label{sec:dr}
According to the underlay spectrum access, small cell transmissions take place on the macrocell spectrum, while satisfying the QoS requirements of the macrocell tier. One way to let the MBS and all the SBSs simultaneously transmit on the same spectral resources is to require that, at each receiver (MUE or SUE) the interfering signal lies on a subspace which is orthogonal to the received useful signal. In this respect, an IA scheme enables the transmitters to achieve high multiplexing gain (or degrees of freedom) by adequately choosing the processing matrices $\textbf{Q}_n$ and $\textbf{V}_k$. By doing so, $\textbf{Q}_n \textbf{H}_{kn} \textbf{V}_k = \textbf{0}$ and $\mathrm{rank}(\textbf{Q}_n \textbf{H}_{0n} \textbf{V}_n)= d_n$ have to be verified by all the MUEs $n$ and the SBSs $k$ \cite{gui2}. The problem of constructing those processing matrices in large multi-tier networks is challenging and the complexity increases when one cannot rely on the coordination between MUEs or SBSs. In fact, this latter case has three important implications. First, the MBS precoding matrix $\textbf{V}_n$ remains fixed regardless of the SBSs' operations. Second, the interference at the MUE's side is generally treated as noise. Finally, due to the contention over the available transmission opportunities, the small cells are limited by the co-tier interference.

In order to apply an IA based solution and benefit from complete interference suppression, the knowledge of the cross channel information $\textbf{H}_{kn}$ is required at each SBS (e.g., it can be acquired assuming channel reciprocity \cite{wu} or through CSI information exchange \cite{6302137}). Furthermore, by considering that the channel coefficients in $\textbf{H}_{kn}$ are identically and independently distributed, the existence of a solution for the IA problem only depends on the dimensions of the problem ($d_n, A_n, d_k, A_k, B$) as discussed in \cite{GT1,yg1}. For example, to let small cell underlaid transmissions fall in the nullspace of the MUEs signal space, the following condition on the number of antennas must be satisfied \cite{nosrat}:

\begin{equation}\label{eq:cond1}
A_k \geq \sum_{n \in \mathcal{N}_{\Phi_k}} d_n + \sum_{j \in \mathcal{K}_{\Phi_k}, \, j \neq k} d_j+  d_k.
\end{equation}

As an example, if each MUE and SUE received one signal stream (i.e., representing one degree of freedom) respectively, the necessary number of transmitting antennas to suppress two interferers would be greater than or equal to three.
When condition (\ref{eq:cond1}) is verified, the small cell deployment reuses the macrocell spatial degrees of freedom and the interference is avoided without modifying the operations at the MUE. This case, known a zero-touch, is of particular interest for heterogeneous networks as discussed in \cite{LTE-A2}. It can be noted that (\ref{eq:cond1}) incurs a limitation on the efficiency of the IA, meaning that the solution exists only for certain properties of the signal (i.e., the number of streams) and the number of antennas equipped at each transmission link. Therefore, when condition (\ref{eq:cond1}) is \emph{not} satisfied, the SBSs can no longer resolve the interference in the spatial domain only. However, each SBS can schedule its transmissions in the spatial and frequency domains, by choosing a spatial precoding strategy and a frequency subchannel. Clearly, by adding the frequency dimension to the problem, the achievable rate depends on the frequency resource management and the scheduling policy at each SBS.
We assume that each SBS $k$ constructs the set $\Phi_k$ by measuring the transmission activity over the macrocell spectrum and selecting the frequency subchannels with the least level of energy. Clearly, due to the nature of the underlay spectrum access, the SBSs compete for the transmission opportunities in space and frequency domains, while, on the other side, the MUEs remain oblivious of the underlaid small cells, and hence non-cooperative.

  \begin{figure}[!t] \vspace*{2cm}
    \centering
       \hspace*{-2cm}\centerline{\psfig{figure=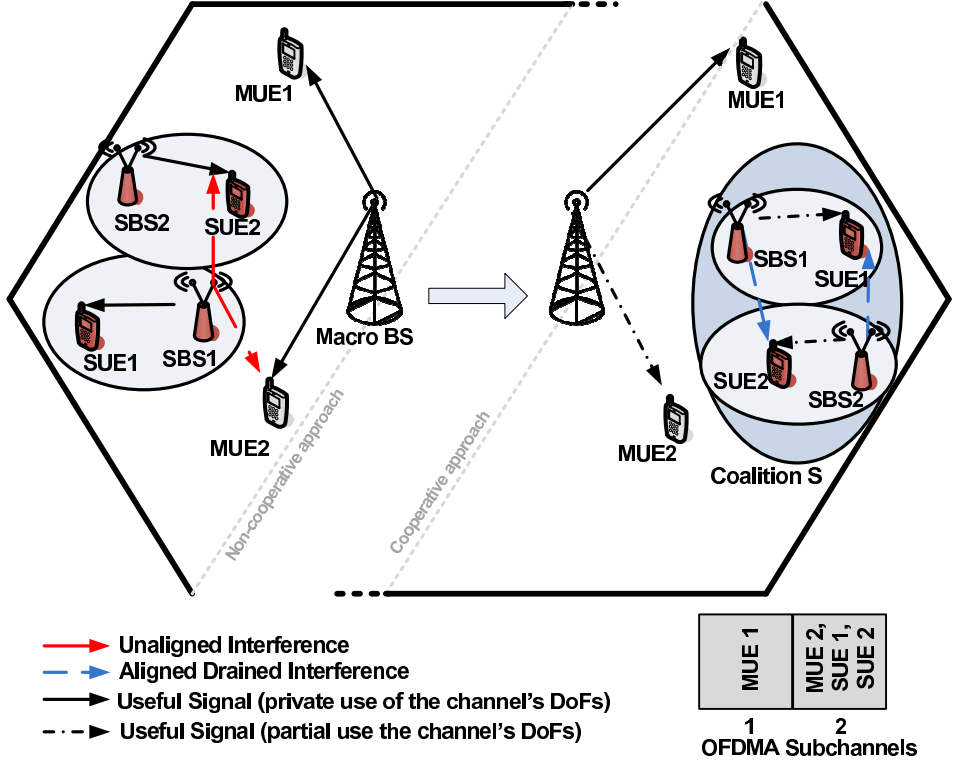,}}
      \caption{A concept model of the proposed solution compared to the traditional non-cooperative approach.}
   \label{fig:scenario}
\end{figure}

Although (\ref{eq:sigf}) includes both co-tier and cross-tier interference contributions, the downlink achievable rate is sensibly limited by the small cell-to-small cell interference, notably when the small cells are densely deployed. To overcome this limitation, we propose an approach using which interfering SBSs decide whether to join cooperative groups, i.e., \emph{coalitions}, and jointly design their precoders so as to reduce the mutual interference. When condition (\ref{eq:cond1}) is verified for all the coalition members, the precoding matrices represent the IA concept solution, which can be obtained through the minimization of the interference leakage, for example \cite{SJ5}. Otherwise, when the macrocell rates decrease due to the small cell transmissions, we propose an interference draining scheme, which generalizes the concept of time-based approach in \cite{drain} to the case of frequency underlaid transmissions. Accordingly, two cooperative SBSs align their transmissions on mutually orthogonal interference subspaces, while maintaining a strong SIR at the MUEs in proximity. Figure~\ref{fig:scenario} illustrates the considered scenario compared to the traditional transmission paradigm.
The conditions for the interference draining can be summarized as follows:

\begin{align}\label{eq:drain}
\begin{aligned}
& \exists \Gamma_S \subset  \mathbb{C}^{B \times d_k}, V_k,V_j \in \Gamma , \\
& \mathrm{span}[\textbf{Q}_i\textbf{H}_{ki}\textbf{V}_k\textbf{s}_k] \perp \mathrm{span}[\textbf{Q}_i\textbf{H}_{ji}\textbf{V}_j\textbf{s}_j] , \forall k,j \in S, \\
&\text{while} \frac{\left \| \textbf{H}_{0n}\textbf{V}_n \textbf{s}_n   \right \|}{\left \| \textbf{H}_{kn}\textbf{V}_k\textbf{s}_k  \right \|} \geq \delta, \forall k \in S,
\end{aligned}
\end{align}\vspace{+0.1cm}

\noindent where $\Gamma_S$ is the interference draining space of coalition $S$ and MUE $n$. Note that, the first condition \label{eq:drain} guarantees that the co-tier interfering components within a coalition $S$ are mutually orthogonal to the useful signal at the respective receivers. The second condition, instead, addresses the interference experienced at the MUEs discovered by the SBSs in $S$ and ensures that a target requirement $\delta$ of signal to interference (SIR) ratio is met. In other words, the precoders $\textbf{V}_k$, $k \in S$ have to verify that the interference brought to the MUEs by the underlay transmissions of the SBSs in $S$ does not excessively deteriorate the MUEs' performance. As an alternative to $\delta$, the impact of the cross-tier interference can also be evaluated by accommodating other metrics, such as the interference constraints \cite{constr1, constr2} or the interference temperature \cite{temperature}. It can be noted that the case of $\left \| \textbf{H}_{kn}\textbf{V}_k\textbf{s}_k  \right \| = 0$ represents the IA solution, since it provides the interference suppression at MUE $n$, through the precoding at SBS $k$. Here, we extend this concept to a coalition $S$ of SBSs, by minimizing the interference leakage caused by the co-channel small cell transmissions, through the precoding matrices $\textbf{V}_k$, $k \in S$. As a result, the problem that we are solving is analogous to construct the precoders so as to solve $\underset{V_k, k \in S}{\arg\min} \left \|\textbf{H}_{kn}\textbf{V}_k \textbf{s}_k \right \|_F$.

We illustrate an example of the proposed interference draining scheme by considering a coalition of two SBSs willing to solve the mutual interference, while respecting the QoS requirements of a nearby MUE. We foresee the following steps:
\begin{enumerate}
\item During the uplink, the cooperative SBSs $j,k$ estimate the channels $\textbf{H}_{nk}$, $\textbf{H}_{nj}$.
\item The cooperative SBSs compute the matrices $\textbf{H}_{kn}$, $\textbf{H}_{jn}$ via channel inversion (assuming channel reciprocity).
\item In the downlink, the SBS $k$, closest to MUE $n$, estimates the subspace spanned by $\textbf{H}_{0n}\textbf{V}_n \textbf{s}_n$ transmissions.
\item At this point, SBSs $j$ and $k$ jointly compute the precoding matrices $\textbf{V}_j$, $\textbf{V}_k$ in the interference draining subspace $\Gamma_k$, i.e.,  that are either in the nullspace of $\textbf{H}_{kn}$,$\textbf{H}_{jn}$, or that verify that the projections of $\textbf{H}_{0n}\textbf{V}_n \textbf{s}_n $ on $ \textbf{H}_{kn} \textbf{V}_k \textbf{s}_k$ and $ \textbf{H}_{jn} \textbf{V}_j \textbf{s}_j$ are greater than $\delta$.
\end{enumerate}

As a result, the interference from members of the same coalition can be suppressed within a coalition $S$, yielding the following signal at SUE  $i\in \mathcal{L}_k$ serviced by SBS $k \in S$:

\begin{equation}\label{eq:ratefc}
\small
R_{i}^{c}=\sum_{d=1}^{d_k} \log \Bigg(1+\frac {\gamma_d^{k} / d_k}{\textbf{e}_d \Big( (\textbf{V}_k^{\dag} \textbf{H}_{ki}^{\dag}\textbf{P}_i \textbf{H}_{ki}\textbf{V}_k  )^{-1} + \sum_{j \in \mathcal{K}_{\Phi_k}\setminus S} \frac{\gamma_d^{j}}{d_j} \textbf{Q}_i \textbf{H}_{ji}\textbf{V}_j \textbf{V}_j^{\dag} \textbf{H}_{ji}^{\dag} \textbf{Q}_i^{\dag}+ \sum_{n \in \mathcal{N}_{\Phi_k}} \frac{\gamma_d^{n}}{d_n} \textbf{Q}_i \textbf{H}_{0i}\textbf{V}_n \textbf{V}_n^{\dag} \textbf{H}_{0i}^{\dag} \textbf{Q}_i^{\dag} \Big) \textbf{e}_d ^{\dag} } \Bigg),
\end{equation}

\noindent where SBS $k$ modulates the $d_k$ signal streams over $|\Phi_k| < d_k$ frequency subchannels, through cooperative interference draining among the SBSs in $S$. Also note how, in (\ref{eq:ratefc}), the residual interference is only imputable to the transmissions from the MBS and the SBSs outside the coalition $S$.

%
%
%


\subsection{Implicit coordination scheme for MUEs and SBSs}
From the small cell perspective, the underlay spectrum access implies that the performance of the MUEs operating over the same spectrum should ideally remain unaffected by transmissions in the other tiers, or at least, that the cross-tier interference remains at a tolerable level. However, it is hard to verify these conditions in absence of coordination among the macrocell and the small cell tiers \cite{wu,FPGC,constr2,noncoop1}.

In a conventional small cell deployment, the rate optimization of the macrocell transmission links is performed by the MBS without accounting on the underlaid small cell transmissions. As a matter of fact, interference-aware rate optimization is a very challenging task in macrocell-small cell networks, mainly because the MBS cannot directly estimate or measure the interference produced by an SBS to an MUE. For example, in order to implement an interference-aware power allocation in the macrocell tier, each SUE is required to measure the interference received from the nearby SBSs, compute the SIR, and convey this information to the MBS. However, It must be stressed that macrocell operations are expected to remain independent of the underlaid small cell deployments. In turn, the small cells are expected to leverage on their self-organizing nature so as to perform the spectrum access and manage the interference.
In line with these considerations, we propose that the MBS performs an autonomous power allocation such as the one proposed in \cite[Section 7.1.1]{tsebook} which does not account for the interference brought by the SBSs, and is only aimed at maximizing the macrocell achievable rate based on the channel realizations. In practice, this means that upon the knowledge of $\textbf{H}_{0n}$, the MBS assigns to the set of signal streams (each one uniquely identified by a frequency subchannel and a spatial direction) a vector $\{P_d^{n*}\} \subset \mathbb{R}^{d_n \times 1}$.

Although the water-filling policy maximizes the achievable rate based on the instantaneous channel condition of each signal stream, it is insensitive of the interference suffered at the MUE. As a result, it could be more rewarding for the MUE to receive the signal streams over the degrees of freedom which are experiencing the best channel conditions and are least affected by the interference. This concept idea, which is also referred to as \emph{modified} water-filling \cite{ioannis}, gives the macrocell tier the flexibility to focus on the degrees of freedom which are more robust to the cross-tier interference. In other words, when a macrocell user is victim of nearby SBSs' transmissions, it can require the MBS to adjust the power allocation and produce a new transmit power vector $\{P_d^{n*}\} \subset \mathbb{R}^{d_n^{*} \times 1}$, which focuses on the least interfered degrees of freedom $d_n^{*}$, while leaving the remaining degrees of freedom unutilized.

Although, the procedure described above appears like an under-utilization of the available resources (i.e., the degrees of freedom), we propose to apply it only when the modified water-filling can compensate the smaller number of degree of freedoms with a higher achievable rate due to the decreased interference. At the same time, in the small cell tier, the newly released resources represent additional transmission opportunities, which can be seized in order to adequately increase the number of degrees of freedom and relieve the congestion during the underlay spectrum access.

As a matter of fact, small cells which incorporate self-organization capabilities are capable of exploiting the unused spatial dimension of the primary link and achieve a throughput improvement, while alleviating the interference on the degrees of freedom currently used by the MUEs. It can be noted that the modified water-filling still requires the MUEs to measure the level of received interference power and to feed it back to the MBS but the mechanism of exploitation of the macrocellular degrees of freedom occurs without involving direct communication between the MUE and the neighboring SBSs. Finally, the achievable rate for MUE $n$ using the modified water-filling policy becomes:

\begin{equation}\label{eq:ratemc}
R_{n}^{c}= \sum_{d=1}^{d_n^{*}} \log \Bigg(1+ \frac {\gamma_d^{n*} / d_n^{*}}{\textbf{e}_d \Big( (\textbf{V}_n^{\dag} \textbf{H}_{0n}^{\dag}\textbf{G}_n \textbf{H}_{0n}\textbf{V}_n  )^{-1} + \sum_{k \in \mathcal{K}_{\Phi_n}} \frac{\gamma_d^{k}}{d_k}  \textbf{Q}_n \textbf{H}_{kn}\textbf{V}_k \textbf{V}_k^{\dag} \textbf{H}_{kn}^{\dag}\textbf{Q}_n^{\dag} \Big) \textbf{e}_d ^{\dag} } \Bigg),
\end{equation}

\noindent where $d_n^{*}$ and $\gamma_d^{n*} = \frac{P_d^{n*}}{\sigma^{2}}$ respectively denote the number of degrees of freedom selected by the modified water-filling policy and the respective signal-to-noise ratio over the d-th stream. Note that, according to such a policy, the rate in (\ref{eq:ratemc}) is achieved over $d_n^* \leq d_n$ degrees of freedom, using transmit power levels $P_d^{n*} \geq P_d^{n}$.

\subsection{Small cell Frequency Reuse}
In order to assess the efficiency of the proposed solution, we provide a comparison with a case in which small cells adopt a frequency reuse scheme. In this case, each SBS senses the macrocell spectrum and modulates its $d_k$-dimensional signal over $|\Phi_k|=d_k$ distinct frequency subchannels (e.g., using an OFDM modulation technique). Clearly, under this approach, the degrees of freedom can only be achieved in the frequency domain, due to the absence of spatial coding. In addition, while the frequency reuse had the advantage of a simpler implementation, since the spectrum access only requires a preliminary sensing phase, it is instead more sensitive to the received interference.
The notion of frequency reuse can be seen as complementary to the IA scheme. In fact, the former allows several transmissions to coexist in the frequency domain while underutilizing the opportunities in the spatial domain. Conversely, the latter exploits the geometrical properties of the received signal to allow the coexistence in the spatial domain, while the frequency dimension is ignored. Intuitively, the interference draining solution, which combines both aspects, can extend the range of operability of the above methods, and thus improve the small cell and macrocell coexistence to further extents. This novel concept has the benefit of solely relying on self-organization capabilities at the small cells, namely in the area of spectrum sensing and dynamic frequency subchannel occupation, which represent the technology requirements of next-generation cellular networks.

\section{Macrocell-small cell coexistence as a coalitional game}
In this section, we analytically model the small cell cooperation framework as a coalitional game in which the MUEs and the SBSs are the players. We introduce some coalitional game concepts and present the solution concept of the recursive core which show the existence of stable coalitions in the networks. Finally, we provide a distributed algorithm which converges to a stable partition.

Let $\Psi = \mathcal{N} \cup \mathcal{K} $ denote the set of the players and in the proposed game and $S \subseteq \Psi$ a coalition in the network, i.e., a set of players which are the decision makers seeking to cooperate. Then, the macrocell-small cell cooperation can be understood as a coalitional game in which, the SUEs form coalitions so as to coordinate the spectrum access and efficiently use the available degrees of freedom in the space and frequency domains, while the MUEs join the existing coalitions to alleviate the received interference.

The overall benefit achieved by a coalition is represented by the \emph{coalitional value} $v(S, \Pi_{\Psi})$, which quantifies the worth of a coalition and is defined as a vector of the individual payoffs of the players in coalition $S$. Further, we recognize that the individual payoff $x_i$ that each player $i$ in coalition $S$ receives is indeed the achievable transmit rate of each SUE and MUE as per $R_{i}^{c}$ and $R_{n}^{c}$ in (\ref{eq:ratefc}) and (\ref{eq:ratemc}), respectively.
According to the coalitional game theory terminology, the game under analysis belongs to the category of coalitional game in \emph{partition form} with \emph{non transferable utility (NTU)}~\cite{WS00,Mac1}. The NTU property is implied by the nature of the transmit rate, which is an individual performance metric that cannot be exchanged among MUEs or SUEs. With respect to the partition form, it must be noted that the value of any coalition $S$ strongly depends on how the players \emph{outside} $S$ have organized themselves, thus, it is affected by the formation of other distinct coalitions in the network. In other words, the performance achieved by each player (SUE or MUE) depends on the partition of the network $\Pi_{\Psi}$ ($\Pi_{\Psi}$ is a partition of $\Psi$). When a coalition is formed, the members jointly remodel their transmit signals in both space and frequency domain, and, to the players outside the coalition, this is seen as a change in the shape of the interference. Therefore, in the proposed model, the rate achieved by the members of any coalition $S \subseteq \Psi$ that forms in the network depends on the cooperative or non-cooperative strategy choice at the SBSs and MUEs in $\Psi \setminus S$.

Now, given two payoff vectors $\textbf{x},\textbf{y} \in\mathbb{R}^{|\Psi|}$, we write $\textbf{x}>_S \textbf{y}$ if $x_i \geq y_i$ for all $i \in S \subset \Psi$ and for at least one $j \in S$ $x_j >y_j$. We also define an \emph{outcome} as couple ($\textbf{x},\Pi_{\Psi}$), where $\textbf{x}$ is a payoff vector resulting from a partition $\Pi_{\Psi}$. Finally, let $\Omega(\Psi,v)$ denote the set of all the possible outcomes of $\Psi$.

\subsection{Recursive core}\label{subsec:rc}\vspace{-0.1cm}

In order to solve the proposed coalition formation game in partition form, we will use the concept of a \emph{recursive core} as introduced in \cite{K01} which is one of the key solution concepts for coalitional games in partition form. In essence, the recursive core is a suitable outcome of a coalition formation process that accounts for externalities across coalitions, which, in the considered game, are represented by the mutual interference between coalitions of SBSs. In order to explain the recursive core we introduce the concept of the \emph{residual game}\cite{K01}.
\begin{definition}
Consider a network $\Psi$ in which a subset of players $S$ has already organized themselves in a certain partition. A \emph{residual game} ($\mathcal{R},v$) is a coalitional game in partition form defined on a set of players $\mathcal{R} = \Psi\setminus S$.
\end{definition}

We use the concept of residual game to model how the rest of the network organizes itself after a coalition $S$ has formed. Clearly, one of the main attractive properties of a residual game is the possibility of dividing any coalitional game in partition form into a number of residual games which, involving a smaller number of players, are easier to solve. Indeed, a residual game is still in partition form and it can be solved as an independent game, regardless of how it was generated. The solution of a residual game $(\mathcal{R},v)$ is known as \emph{the residual core} which is defined as the set of possible game outcomes, i.e., partitions of $\mathcal{R}$ that can be formed.

\noindent Through the concept of residual games, it is possible to analyze the cooperative behavior in large networks in a computationally easier way as the residual games are defined over a smaller set of players than the original game. Hence, the recursive core solution can be found by recursively playing residual games, which yields the following definition\cite[Definition~2]{K01}:

\begin{definition}\label{def:Reccore}
The \emph{recursive core} of a coalitional game $(\Psi,v)$ is inductively defined in four main steps:

\begin{enumerate}
  \item \emph{Trivial Partition}. In a network with only one player $\Psi\!\!=\!\!\left \{i \right \}$, the recursive core is clearly composed by the only outcome with the trivial partition composed by the single player $i$: $C(\left \{ i \right \}, v )$ = $(v({i}),i) $.
  \item \emph{Inductive Assumption}. As an inductive step, we assume that the residual games $(\mathcal{R},v)$ with at most $K-1$ players have been defined and each one is associated to a residual core $C(\mathcal{R},v)$. Thus, proceeding recursively, we define the \emph{assumption} $A(\mathcal{R},v)$ about the game $(\mathcal{R},v)$ as follows:  $A(\mathcal{R},v)=C(\mathcal{R},v)$, if $C(\mathcal{R},v) \neq \varnothing$ ; $A(\mathcal{R},v)=\Omega(\mathcal{R},v)$, otherwise. In other words, an assumption defines a preference on how to partition a residual game $\mathcal{R}$, and it coincides with the residual core, if already defined, or with the set of any possible partition of $\mathcal{R}$.
  \item \emph{Dominance}. We now introduce the mechanisms of selection among the possible partitions. An outcome $(\textbf{x},\Pi_{\Psi})$ is \emph{dominated} via a coalition $S$ if for at least one $(\textbf{y}_{\Psi\setminus S},\Pi_{\Psi\setminus S}) \in A(\Psi\setminus S,v)$ there exists an outcome $((\textbf{y}_S , \textbf{y}_{\Psi\setminus S}), \Pi_S \cup \Pi_{\Psi \setminus S}) \in \Omega(\Psi,v)$ such that $(\textbf{y}_S , \textbf{y}_{\Psi\setminus S})>_S \textbf{x}$.
  \item \emph{Core Generation}. Finally, the recursive core of a game of $\left | \Psi \right |$ players is the set of undominated outcomes and we denote it by $C(\Psi,v)$.
\end{enumerate}

\end{definition}

One can notice that a stable network partition will emerge according the concept of dominance in step 3) of Definition \ref{def:Reccore}. The concept of dominance inherently captures the fact that the value of each coalition depends on the belonging partition. Hence, it can be expressed in the following way. Given a current partition $\Pi_{\Psi}$ and the associated payoff vector $\textbf{x}$, an undominated coalition $S$ represents a deviation from $\Pi_{\Psi}$ such that the resulting outcome $((\textbf{y}_S, \textbf{y}_{\Psi\setminus S}), \Pi_S \cup \Pi_{\Psi\setminus S})$ is more rewarding for the players of $S$.
It appears clear now that by simultaneously playing reduced games, the players organize themselves in the coalitions which guarantee the highest payoff, which is uniquely determined by the belonging partition. Thus, finally, the recursive core can be interpreted as the set of those undominated partitions.

We now analyze the stability of the recursive core solution and provide some instructions in order to guarantee it.
As the players of the game under analysis are MUE and SUE which face different operation as described in Section III, the stability of the partition in the recursive core has to verify diverse conditions. With respect to the small cell tier, it can be observed that the dominant interferers SBSs are the most eligible to join a SBSs coalition. Their respective transmit rates are limited by the received signals overlapping in the frequency and spatial dimension, thus, through cooperation they would jointly construct the precoders in order to suppress the mutual interference. As a result, as long the coalitions are constructed while iteratively suppressing the interference among dominant interferers, the members will not abandon it and the coalition value will be non decreasing at each iteration.
At the MUE side, when a cooperative strategy is adopted the MUE is associated with the coalition exploiting its unused degrees of freedom, although there is no direct interaction with the observed MUE and the SBSs in the coalition. Naturally, as an MUE may release its degrees of freedom only upon a feasible reduction of the received interference, the transmit rate achieved by MUE $n$ in coalition $S$ over the degrees of freedom $d_n^{*}$ has to verify the condition: $R_{n}^{c} \geq R_{n}$.

\bigskip
\subsection{Proposed Algorithm and Distributed Implementation}
In the following we provide a distributed algorithm which converges to the recursive core and reflects the above considerations on how stable coalition form.

\begin{algorithm}[!t]\label{ALG:core}
  \scriptsize
  \flushleft
      \caption{{\small Distributed coalition formation algorithm for interference draining in small cell networks}}
      \label{ALG:recursivecore}
      \begin{algorithmic}
          \STATE{\textbf{Initial State at the SBS:} The network is partitioned by $\Pi_{\Psi}$ = $\mathcal{N} \cup \mathcal{K}$ with non-cooperative SBSs and MUEs.}
          \STATE{\textbf{Proposed Coalition Formation Algorithm}}
          \STATE{\emph{Phase I - Interferers Discovery}}
          \STATE\hspace*{1em}{1) Based on the collected RSSIs, each SBS $k$ discovers the interfering SBSs $j$, and forms an\\ \hspace*{2.2em}interferers list sorted by the level of interference brought to the SUEs $i \in \mathcal{L}_k$.}
          \STATE\hspace*{1em}{2) During the UL, each SBS $k$ estimates the subspace spanned by $\textbf{H}_{nk}\textbf{V}_n$ from MUEs\\ \hspace*{2.2em}transmissions and identifies an interference draining subspace $\Gamma_k$.}\vspace*{1em}
          \STATE{\emph{Phase II - Small Cell Coalition Formation}}
          \FORALL{SBS $j$ in the list}
          \STATE{1) SBS $k$ computes a precoding matrix $\textbf{V}_{k} \in \Gamma_k$ which guarantees the first draining condition in (6)\\ \hspace*{1.2em} for all the SUEs $l\in \mathcal{L}_j$.}
          \STATE{2) Each SBS $k$ computes the projection of $\textbf{s}_k$ on the signal subspace of each of the detected MUEs\\ \hspace*{1.2em} $n$, and computes the respective SIR.}\vspace*{0.5em}
          \IF{$\textbf{V}_{k}$ verifies the second condition in (6)}
                      \STATE{3) SBS $k$ sequentially engages in pairwise negotiations with SBS $j$ in the list to join coalition $S$.}
                      \STATE{4) Each SBS evaluates the average rate $R_i^{c}$ of its SUE $i$ as in (\ref{eq:ratef}).}            \ELSE
                      \STATE{5) Current SBS $j$ is discarded and the following SBS in the interferers list is assessed.}
                      \ENDIF\vspace*{0.5em}
                      \STATE{6) The payoff is updated, accounting for the newly adopted strategy.}
                      \STATE{7) Each SBS joins the SBS which ensures the maximum payoff.}
                      \ENDFOR\vspace*{0.5em}

          \STATE{\textbf{Outcome of this phase:} Convergence to a stable partition $\Pi_{\Psi}$ in the recursive core.}\vspace*{1em}
          \STATE{\emph{Phase III - Coalition-level CoMP transmissions}}
          \STATE\hspace*{1em}{1) Within each coalition, cooperative interference draining operations as described in Section \ref{sec:dr} are\\ \hspace*{2.2em}initiated.}\vspace*{0.5em}

          \STATE{\textbf{Initial State at the MUE:} Each MUE $n$ controls the SIR over each of the $d_n$ signal streams.}
          \IF{the interference on the d-th signal leads to a SIR smaller than $\frac{\delta}{d_n}$}
          \STATE{MUE $n$ executes the modified water-filling algorithm and updates the rate $R_n^{c}$.}
          \IF{$R_n^{c}\geq R_n$ }
          \STATE{The d-th degree of freedom is released, and the payoff update to $R_n^{c}$.}
          \ENDIF
          \ENDIF\vspace*{0.5em}
      \end{algorithmic}
\end{algorithm}\vspace{-0.1cm}

To reach a partition in the recursive core, the players in $\Psi$ use Algorithm~\ref{ALG:recursivecore}. In this algorithm, which includes the operations at both the SBS and the MUE sides, we devise three phases: Interferer discovery, small cell coalition formation, and coalition-level cooperative transmission. Initially, the network is partitioned by $ \left |\Psi\right |$ singleton coalitions (i.e., non-cooperating mobile users).
During the phase of interference discovery, the MBS periodically requests Received Signal Strength Indicators~(RSSIs) measurements from its MUEs to identify the presence of small cells which might cooperatively provide higher throughput. Then, based on the RSSIs, the interfering SBSs are ordered from the stronger to the weaker. Moreover, during the uplink~(UL) macrocell transmissions, each SBS estimates the subspace spanned by $\textbf{H}_{nk}\textbf{V}_n$ of any MUE $n$ in proximity. This operation is accompanied by the assumption of reciprocity of channel $\textbf{H}_{nk}$, to allow SBS $k$ to estimate the interference produced at the MUE $n$.
In the successive phase of coalition formation, each SBS selects the first interfering SBS from the ordered list and computes the  precoding matrix $\textbf{V}_k$ which verifies the first condition in (6). If also the second condition in (6) is verified for all the MUEs detected by the negotiating SBSs, SBS $k$ sends a request for cooperation to its counterpart. If both SBSs mutually approve the cooperation request, they form a coalition $S$, and their transmissions will lay in the interference draining space $\Gamma_S$. Once a coalition has formed, the member SBSs exchange information for properly model the matrices $\textbf{Q}_k$, $\textbf{V}_k$ which realize the draining of the interference, and the channel state indicators $\textbf{H}_{ji},\textbf{H}_{ki}$, via the X2 interface\footnote{Nevertheless, the data exchange among neighboring SBSs can also occur via wireless link  or through the wired backhaul.}.

For the MUEs, we assume that no direct cooperation with the small cell tier occurs, however, each MUE can estimate the SIR of each of the $d_n$ received signal streams. Therefore, when the SIR level at the generic stream $d$ is lower than the average threshold $\frac{\delta}{d_n}$, and the modified water-filling policy over the remaining degrees of freedom guarantees a higher rate, then the d-th degree of freedom is released by allocating the power over the remaining stream. As the SBSs are able to detect the dimensions of the MUE signal subspace, the newly released degrees of freedom have a beneficial impact in finding a solution which respects the QoS requirements as per (6). Further, this results in a more efficient interference management of the SBSs which, in return, spread the transmit power over a larger number of streams, and flatten the interference over a larger set of degrees of freedom.

Next, we prove the following property for our algorithm:
\begin{property}\label{prop:conv}
Using Algorithm~\ref{ALG:recursivecore}, coalitions of SBSs merge together by Pareto dominance, and, thus, the resulting network partition $\Pi_{\Psi}$ is \emph{stable and lies in the recursive core} $C(\Psi,v)$ of the game.
\end{property}

\noindent \emph{Proof.} Each distributed decision taken by an SBS defines the shape of a coalition in the network, hence, the shape of the overall network partition. Therefore, Algorithm~\ref{ALG:recursivecore} can be seen as a sequence of steps through which the SBSs sequentially transform the composition of the network partition.
For example, let us assume that the network at a given step $t$ is partitioned by $\Pi_{\Psi}^{(t)}$, and that an SBS $k \in S \subset \Pi_{\Psi}^{(t)}$ deviates to another coalition $T \subset \Pi_{\Psi}^{(t+1)}$, which Pareto dominates $S$. In other words, if $\textbf{x}$ and $\textbf{y}$ are the payoffs vectors of coalitions $S$ and $T$, respectively, $x_k < y_k$ and $x_j \leq y_j$ for all $j \in T \subset \Pi_{\Psi}^{(t+1)}$. Note that, as each SBS gradually selects the partners among its mutual interferers without affecting the other orthogonally allocated SBSs in the network, the value of other coalitions remains unchanged.
Therefore, given any two successive algorithm steps $t$ and $u$, $t<u$, we have that $\Pi_{\Psi}^{(t)}$ is Pareto dominated by $\Pi_{\Psi}^{(u)}$.
As a result, $ v(\Pi_{\Psi}^{(t)}) = \sum_{S \in \Pi_{\Psi}^{(t)}} v(S,\Pi_{\Psi}^{(t)})  <  v(\Pi_{\Psi}^{(u)}) = \sum_{T \in \Pi_{\Psi}^{(u)}} v(T,\Pi_{\Psi}^{(u)})$.

The above sequence resulting from the proposed algorithm ensures that the overall network utility sequentially increases by Pareto dominance. Thus, at each iteration of Algorithm~\ref{ALG:recursivecore}, the sum of values of the coalitions in the network increases without decreasing the payoffs of the individual SBSs. We show that as the number of possible steps of the algorithm is finite and given by the number of possible partitions of $\Psi$ (Bell number \cite{Mac1}), Algorithm~\ref{ALG:recursivecore} converges to a final partition.

When an SBS cannot find any other deviation which is profitable by Pareto dominance, it has reached the highest payoff and then induced an undominated coalition which lies in the recursive core of the game.  Clearly, the players have no incentive to deviate from the current partition, because any other possible strategy would lead to an inferior payoff. The partition in the recursive core is therefore stable since, once formed, it will not change into any other partition provided that the players are always able to modify their strategy at any time. $ \:\:\:\:\:\:\:\: \:\:\:\:\:\:\: \:\:\:\:\:\:\:\: \:\:\:\:\:\:\:\: \:\:\:\:\:\:\:\: \:\:\:\:\:\:\:\: \:\:\:\:\:\:\:\: \:\:\:\:\:\:\:\: \:\:\:\:\:\:\:\: \:\:\:\:\:\:\:\: \:\:\:\:\:\:\:\:  \:\:\:\:\:\:\: \:\:\:\:\:\:\:\: \:\:\:\:\:\:\:\:  \:\:\:\:\:\:\: \:\:\:\:\:\:\:\:  \:\:\:\:\:\:\: \:\:\:\:\:\:\:\: \:\:\:\:\:\:\:\:\:\:\:\:\:\:\:\:\:\: \:\:\:\:\:\:\:\: \blacksquare$

\smallskip

Therefore, the recursive core is reached by considering that only the payoff-maximizing coalitions are formed, through the concept of dominance in  Definition \ref{def:Reccore}. Clearly, this algorithm is distributed since the SBSs and MUEs take individual decisions to join or leave a coalition, while, ultimately reaching a stable partition, i.e., a partition where players have no incentive to leave the belonging coalition. Those stable coalitions are in the recursive core at the end of the second stage of the algorithm. Finally, once the coalitions have formed, the members of each coalition proceed to perform the interference draining operations described in Section \ref{sec:dr}.

\section{Numerical Results}
\noindent For system-level simulations, we consider a single macrocell with a radius of $650$~m within which $K$ SBSs and $N$ MUEs are randomly distributed. Each SBS $k \in \mathcal{K}$ serves $L_k=1$ SUE scheduled over $|\Phi_k|$ subchannel, adopting a closed access policy. We set the maximum transmit power per transmission at the MBS and the SBSs to $P_{max}^{n}=40$~dBm, $P_{max}^{k}=20$~dBm, respectively. Transmissions are affected by distance dependent path loss shadowing according to the 3GPP specifications \cite{3GPP3}. Moreover, a wall loss attenuation of $12$~dB affects SBS-to-MUE transmissions. The considered macrocell has $200$ available subchannels, each one having a bandwidth of $180$~KHz. The MBS and each SBS dedicate $|\Phi_n|=1$ and $|\Phi_k| \leq 4$ subchannels to the transmission of each MUE and SUE, respectively. For both SUEs and MUEs, we assume that power control fully compensates for the path loss. Further simulation parameters are included in Table 1. To leverage channel variations and user positions, statistical results are averaged on a large number of simulation rounds (Monte Carlo simulations).

\begin{table}[ht]
\scriptsize
\caption{Table 1 - Small cell System Parameters}
\centering
      \begin{tabular}{|c|c||c|c|}
       \hline
  Macrocell radius & 650m & Number of antennas at the MUE, SUE   & $B = 2$ \\
  Small cell radius  & 15-25m & Max TX power at MBS (SBS): $P_{max}^{n}$ ($P_{max}^{k}$)  &  40 dBm (20 dBm) \\
  Carrier frequency & 2.0 GHz & Forbidden drop radius (SBS) & 0.2m \\
  Number of SBSs   &  1 - 360  & Total Bandwidth & 40 MHz\\
  Number of SUEs per small cell ($L_k$)  &  1  & Subcarrier Bandwidth  & 180 kHz \\
  Number of MUEs per macrocell  & 1- 200 & Thermal Noise Density &  -174 dBm/Hz \\
  Minimum required SIR at each MUE: $\delta$  & 8-12 dB  & Path Loss Model [dB] (outdoor)& $15.3+ 37.6\log_{10}$(d[m])\\
  SBS antenna gain  &  0 dBi & External wall penetration loss &  12dB\\
  Forbidden drop radius (macro) & 50m & Lognormal shadowing st. deviation  & 10 dB  \\
  Number of antennas at the MBS (SBS) &  $A_n =\{2,4\}$ ($A_k =\{2,4\}$) & Shadowing  correlation between SBSs  & 0  \\
  \hline
      \end{tabular}
\end{table}

\begin{figure}[!t]
    \centering
       \centerline{\psfig{figure=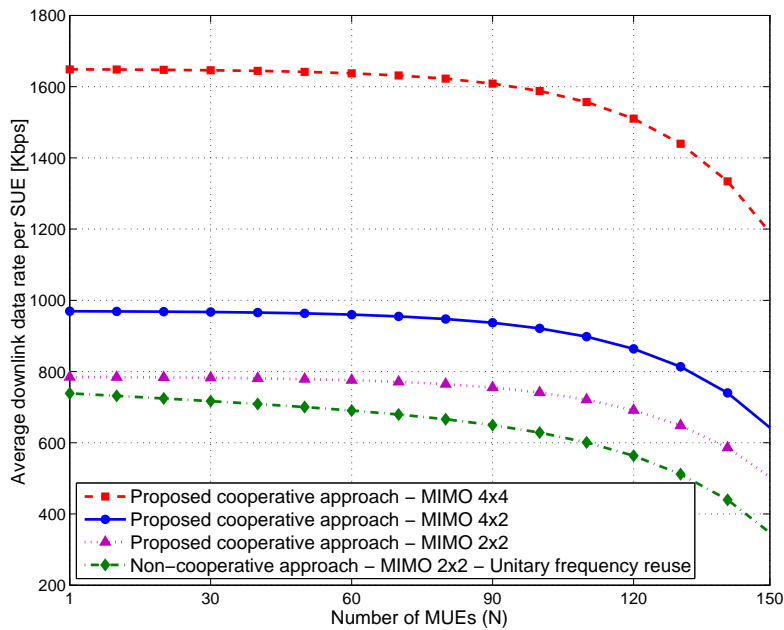, width=12cm}}
   \caption{Average individual payoff per SUE as a function of the number of MUEs, for the different studied approaches and MIMO antenna sets. $\delta=12$~dB, $K=200$.}
   \label{fig:payoff_SUE}
\end{figure}

In Figure~\ref{fig:payoff_SUE}, we show the average payoff per SUE as a function of the number of MUEs in the network $N$, for different strategies and MIMO antenna set sizes $A_k \times B =\{2,4\} \times \{2,4\}$. Figure~\ref{fig:payoff_SUE} shows that a cooperative strategy whose solution is based on the joint interference draining leads to gains almost proportional to $B$. Nevertheless, as the density of MUEs grows, the average rates start decreasing as the mechanism of interference suppression approaches the maximum number of signals which can be suppressed. For instance, Figure~\ref{fig:payoff_SUE} shows that the average payoff per SUE with a 4x2 MIMO antenna set resulting from the coalition formation can achieve an additional $51\%$ gain with respect to the non-cooperative case, in a network with $K=200$ SBSs and $N=120$ MUEs. Therefore, we demonstrated how cooperation is beneficial to the SUEs in highly populated areas where the density of interferers (i.e., potential coalitional partners) is high.


\begin{figure}[!t]
    \centering
       \centerline{\psfig{figure=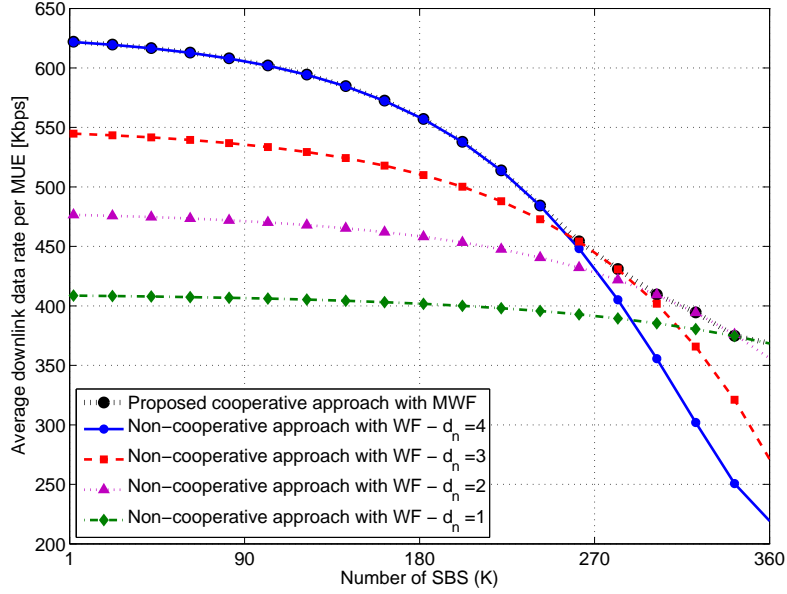, width=12cm}}
   \caption{Average individual payoff per MUE as a function of the number of MUEs, for the different studied approaches and MIMO antenna sets. $\delta=12$~dB, $N=150$.}
   \label{fig:payoff_MUE}
\end{figure}

In Figure~\ref{fig:payoff_MUE}, we evaluate the performance of the proposed coalition formation game model by plotting the average payoff achieved per MUE during the whole transmission time scale as a function of the number of SBSs $K$. We compare the performance of the proposed algorithm to that of the non-cooperative case, for different number of signal streams per MUE $d_n= 1 - 4$. It can be noted that the MUE achievable rate is affected by the cross-tier interference in a way which is proportional to the portion of spectrum occupied. As the number of SBSs further grows, the interference brought at the MUE side justifies a cooperative approach with modified water-filling power allocation, as it grants a larger achievable rate. Hence, the MUEs will successively release the available degrees of freedom while perceiving a reduction on the received interference. For example, Figure~\ref{fig:payoff_MUE} shows that by releasing $2$ degrees of freedom, an MUE can gain up to $33\%$ with respect than the non-cooperative case in a network with $K=320$ SBSs and $N=150$ MUEs.

\begin{figure}[!t]
    \centering
       \centerline{\psfig{figure=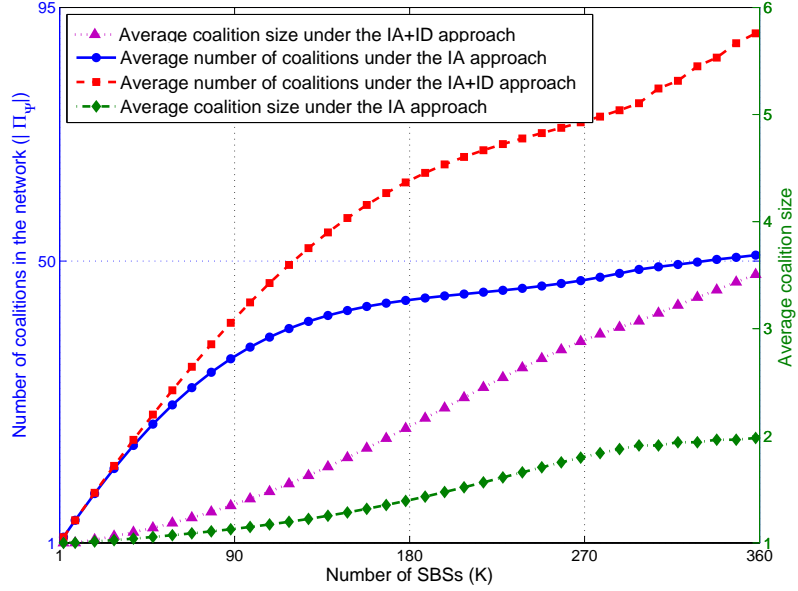, width=12cm}}
   \caption{Average coalition size as function of the number of SBSs, for different QoS requirements at the MUEs, expressed by $\delta=12$~dB and $\delta=8$~dB in the ID+IA and IA approaches respectively. $N=200$.}
   \label{fig:coalsize}
\end{figure}

In Figure~\ref{fig:coalsize}, we observe the average number of coalitions in the network and the average size of the SBS coalitions in the recursive core for a given QoS target of $\delta=12$~dB at each MUE. Figure~\ref{fig:coalsize} shows that, for small networks, $K < 40$~SBSs, the SBSs have low incentive to cooperate, and, thus, the recursive core is mainly populated by singleton coalitions. Conversely, for larger network sizes ($40<K<160$~SBSs), the SBSs start to prefer a cooperative strategy, as witnessed by the increase in the average size of the coalitions. The coalition formation becomes even more preferable when the SBSs can exploit the frequency dimension as it extends the limitation of condition (\ref{eq:cond1}). Indeed, by doing so, nearby SBSs can drain the mutual interference on signal subspaces, which are mutually orthogonal among the coalition members and still respect the QoS requirement $\delta$ at the MUE close to any of the coalition members. Further, for $K>160$~SBSs, also note how the IA based approach cannot accommodate new coalition members as the solution reaches a saturation point, while the interference draining allows for additional gains reaching up to an average coalition size of $3$ for a network with $K=280$ SBSs, with respect to the $1.8$ of the IA based approach.


\begin{figure}[!t]
    \centering
       \centerline{\psfig{figure=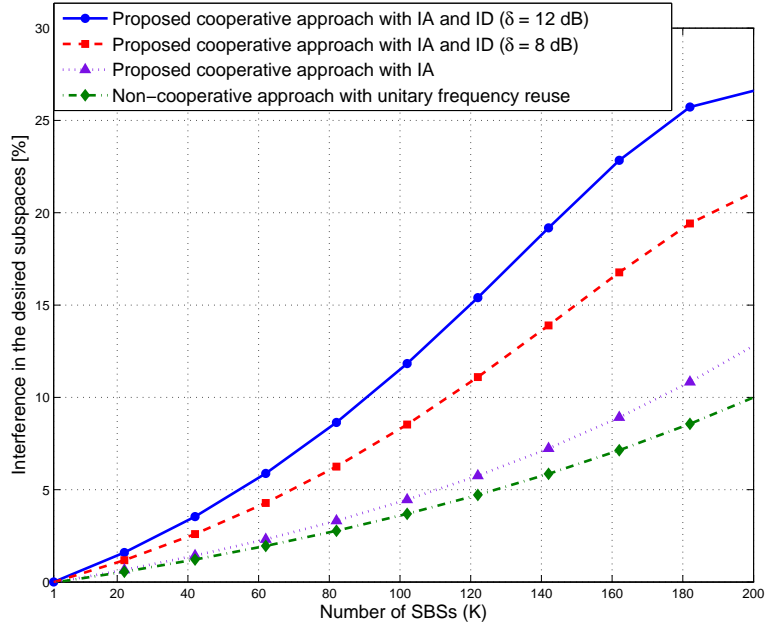, width=12cm}}
   \caption{Average percentage of interference in the desired signal subspace under different approaches versus the total number of SBSs. $\delta= 12 dB$.}
   \label{fig:drainn}
\end{figure}

Figure~\ref{fig:drainn} shows the efficiency of the proposed solution in terms of percentage of interference in the desired signal signal subspace versus the number of SBSs in the network. In this figure, we show that through cooperative operations it is possible to redirect the interference over signal subspaces which are mutually orthogonal among coalition members. In a non-cooperative approach, the interference is randomly distributed over the spectrum channels and the spatial directions, so the ascendant behavior in Figure~\ref{fig:drainn} is a consequence of the number of transmissions which linearly grows with $K$. Conversely, through the proposed approach with interference draining it is possible to control the addressed interference subspace and this allows for additional interference reduction of $17\%$ with respect to the non-cooperative case. As the number of SBS gets larger ($160<K<200$), the spectrum becomes congested and the interference starts to occupy all the signal subspaces (i.d., the degrees of freedom) in the network, with a consequent impact decrease on the achievable gains. Finally, note how the benefit of the proposed cooperative approach grows in case of a higher tolerance of the MUEs' to the received interference.

\begin{figure}[!t]
    \centering
       \centerline{\psfig{figure=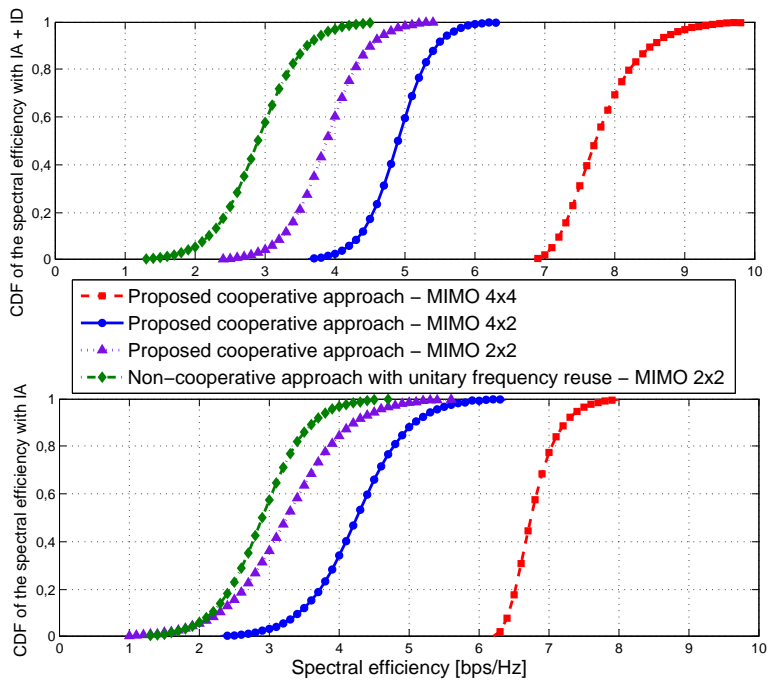, width=12cm}}
   \caption{Cumulative distribution function of the spectral efficiency per SBS for different studied approaches and MIMO antenna sets.  $\delta = 12$~dB. $N=200$, $K=200$.}
   \label{fig:cdf}
\end{figure}

In Figure~\ref{fig:cdf}, we compute the cumulative distribution function of spectral efficiency of the proposed approach for different number of antennas $A_k \times B = 4\times4, 4\times2, 2\times2$, in a network with $N=250$~MUEs and $K=250$~SBSs. This figure shows that through spatial reuse it is possible to significantly reduce the co-tier interference and achieve high spectral efficiencies. In detail, we compared a solution which is only based on the interference alignment with one that performs the interference draining in the spatial and frequency domains. It can be noted that the proposed interference draining solution results in a further improvement of $15\%$ of the average spectral efficiency per small cell transmission, with respect to the IA solution. This is motivated by the fact that, when only an IA based solution is available, the coalition formation process reaches its saturation for smaller network sizes. Therefore, through under an IA based approach it is possible to form coalitions and solve less interfering links than under an interference draining approach.

\begin{figure}[!t]
    \centering
       \centerline{\psfig{figure=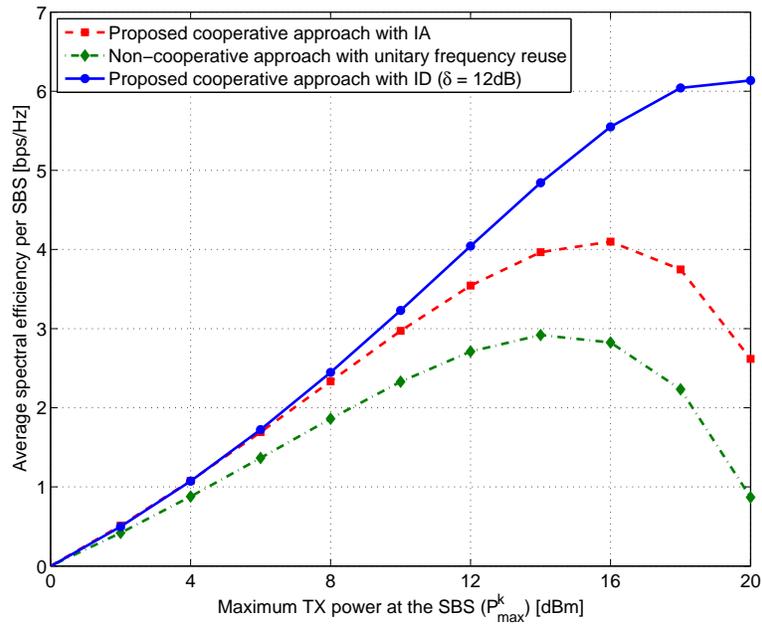, width=12cm}}
   \caption{Average spectral efficiency per SBS link vs maximum transmit power limits, for the different studied approaches.  $\delta = 12$~dB. $N=200$, $K=250$ $A_k=4, B =2$.}
   \label{fig:sef}
\end{figure}

Figure~\ref{fig:sef} shows the average spectral efficiency per small cell link as a function of the maximum transmit power $P_{max}^{k}$ at each SBS, for different studied approaches, in a network with $N=150$ SBSs. For low levels of transmit power $P_{max}^{k} < 6$~dBm, the performance of the ID and IA based approaches are similar, as the interference among SBS is limited. As the level of transmit power increases ($6 <P_{max}^{k} < 16$~dBm), the mechanisms of interference avoidance outperform the traditional non-cooperative frequency reuse scheme. Furthermore, it can be observed that the ID based approach allows for a more effective interference management, for higher transmit power levels, when the received interference is generally the main factor of low SIRs. In fact, we observe that cooperative SBSs using an ID based approach can gain up to $35\%$ and $89\%$ with respect to an IA based approach and a non-cooperative case, respectively. Finally, for $P_{max}^{k} > 16$~dBm, the average spectral efficiency gains eventually decrease, being limited by the co-tier interference. In a nutshell, Figure~\ref{fig:sef} demonstrates that the proposed coalitional game model using interference draining has a significant advantage over the non-cooperative case, which increases with the MUEs' toleration to the cross-tier interference.

\section{Conclusions}

\noindent In this paper, we have introduced a cooperative framework for interference mitigation in both the small cell and the macrocell tiers. We have formulated the problem as a coalitional game in partition form and proposed a distributed coalition formation algorithm that enables SBSs to independently select the most rewarding strategy, while accounting for a limitation on the interference brought to the close MUEs. We have shown that the proposed algorithm reaches a stable partition, which lie in the recursive core of the studied game. Within every formed coalition, we have proposed an interference draining scheme, which is a suitable solution for enabling multiple underlay transmissions over the same spectrum. Results have shown that the performance of underlay small cells is ultimately limited by the received interference, therefore, the proposed cooperative strategy among interfering small cells brings significant gains, in terms of average achievable rate per small cell, reaching up to $37\%$, relative to the non-cooperative case, for a network with $150$ MUEs and $200$ SBSs.



\newpage
\bibliographystyle{IEEEtran}
\bibliography{references}






\end{document}